\documentclass[pra,twocolumn,showpacs,preprintnumbers]{revtex4}
\usepackage{graphicx}
\usepackage{dcolumn}
\usepackage{bm}
\usepackage{amsmath}

\begin{document}

\title{Exact results of two-component ultra-cold Fermi gas in a hard wall trap}
\author{Bo-Bo Wei$^1$}
\author{Jun-Peng Cao$^{1,2}$}
\author{Shi-Jian Gu$^1$}
\author{Hai-Qing Lin$^1$}

\affiliation { $^{1}$ {Department of Physics and Institute of
Theoretical Physics, The Chinese University of Hong Kong, Hong Kong,
China} \\ $^{2}$ {Beijing National Laboratory for Condensed Matter
Physics, Institute of Physics, Chinese Academy of Sciences, Beijing
100190, China} }

\begin{abstract}

The ground state properties of a one-dimensional two-component
ultra-cold Fermi gas in an infinite potential well are investigated.
The wave function of the system is obtained explicitly based on the
solution of Bethe ansatz equations. The single-particle reduced
density matrix and two-particle density-density correlations are
also evaluated. It is found that the momentum density distributions
of the strongly interacting two-component Fermi gas is significantly
different from that of free spinless Fermi gas while the density
distributions in real space are similar.

\end{abstract}

\pacs{05.30.Fk, 67.85.-d, 71.10.-w}
\date{\today}
\maketitle




\section{Introduction}

There have been remarkable experimental advancements in the area of
trapped one-dimensional cold atom systems
\cite{AGorlitz,HMoritz,TKinoshita,BLTolra}.  The one-dimensional
quantum gas can be realized experimentally by tightly confining the
atomic cloud in the radial directions and weakly confining it along
the axial direction \cite{RecatiA,Tokatly}. Since the radial degrees
of freedom are frozen \cite{MOlshanii,BEGranger}, the quantum gas
can be effectively described by a one-dimensional model along the
axial direction.  Using Feshbach resonance, the energy of a bound
state of two colliding atoms in a magnetic field was tuned to vary
the scattering length from $-\infty$ to $\infty$, which allows
experimental access to strongly interacting regime and studying the
crossover from a BCS to Bose-Einstein Condensation.
\cite{DSPetrov,VDunjko,POhberg,T.Bergeman}. The experimental
achievements have opened up many exciting possibilities for
theoretical investigation of quantum effects in low-dimensional
many-body systems.

Exactly solvable models provide us an important insight in studying
the quantum many-body physics beyond various approximation schemes.
The exact solutions can supply some believable results thus serve as
a very good starting point to understand the new phenomena and new
quantum states in trapped cold atomic systems. Recently, the study
of one-dimensional bosons with contact interactions in an infinite
potential well shows a revival of interest \cite{HaoA,MTBatchelor}.
We note that the point-interacting two-component Fermi gas in a hard
wall trap is also interesting. Many new physics appears in the
fermionic system and most important is that this system is
integrable \cite{MGaudin,CNYang,F.Woynarovich,NOelkers,Tujimotok}.
Meanwhile, this systems can be realized in experiment by the laser
cooling technique.

In this paper, we investigate the one-dimensional ultra-cold
two-component Fermi gas in an infinite potential well. Most of the
previous studies focus on the energy spectrum and its related
quantities. Here, we put our attentions on the wave function of
the system. We study the single-particle reduced density matrix,
natural orbitals, the momentum density distributions and two-body
correlations based on the exact wave function of the system. These
quantities are important for a better understanding of the ground
state properties of the fermion system and can be detected in
experiments directly.

The paper is organized as follows. In Sec.~II, the model Hamiltonian
and its Bethe-ansatz solution are introduced in details. In
Sec.~III, we study the one-body aspects of correlations.
Specifically, we present single-particle reduced density matrix in
the context of long range order in Sec.IIIA. In Sec.~IIIB, We
discuss the natural orbitals and their populations. The momentum
density distributions are shown in Sec.~IIIC. In Sec.~IV, we go
further to study the two particle correlations. A summary of our
main results is given in Sec.~V.

\section{THE MODEL AND ITS EXACT SOLUTION}

\label{sec:res}

We consider a system of two-component Fermi gas with $\delta$
interaction confined in one-dimensional box of length $L$. The
model Hamiltonian reads
\begin{equation}
{\cal H}=-\sum_{i=1}^N{\frac{\partial^2}{\partial x_i}}
+2c\sum_{i<j}^{N}\delta(x_i-x_j),
\end{equation}
where $c$ is the interaction strength between different spins of
fermions and $N$ is total number of fermions. The coupling $c$ can
be expressed in terms of the scattering length as $c=2/a_{1D}$,
where $a_{1D}$ is the effective one-dimensional scattering length.
The $c=0$ and $c\rightarrow \infty$ limits correspond to zero and
infinite interactions, respectively. It can be tuned by the
scattering length. The coupling is repulsive for $c>0$ and
attractive for $c<0$. We only consider the repulsive case in this
work. We introduce another important parameter $\gamma$ that
characterizes the different physical regimes of one-dimensional
quantum gas, $\gamma=c/ \rho$ where $\rho=N/L$ denotes the linear
density of the system.

The wave function of system (1) is antisymmetric, and can be
characterized by the total number of fermions $N$ and the number
of spin-down fermions $M$. Suppose the antisymmetric wave function
is $\Psi(x_1\sigma_1,x_2\sigma_2,\cdots,x_N\sigma_N)$, where $x_j$
and $\sigma_j$ are the position coordinates and spin coordinates
of $j$-th fermion, respectively. Then the static Schr\"{o}dinger
equation for the wave function $\Psi$ reads \cite{E. Lieb}:
\begin{eqnarray}
\left(-\sum_{i=1}^N{\frac{\partial^2}{\partial x_i}} +2c\sum_{1\leq
i<j\leq N}\delta(x_i-x_j)\right)\Psi=E\Psi.
\end{eqnarray}
The explicit form of the wave function of the system in  region
$x_{Q_1}\leq x_{Q_2}\leq \ldots \leq x_{Q_N}$ reads
\cite{F.Woynarovich}
\begin{eqnarray}
& &\Psi(x_1\sigma_1,x_2\sigma_2,\cdots,x_N\sigma_N)
\nonumber\\
&=& \sum_{P,r_1,\dots,r_N}(-1)^P (-1)^Q
A_{\sigma_{Q_1},\ldots,\sigma_{Q_N}}(r_1 k_{p_1},\ldots,r_N
k_{p_N})\nonumber \\ & & \times \exp(i\sum_j r_j k_{p_j}x_j).
\end{eqnarray}
Here $P$ and $Q$ are permutations of $1,2,\ldots,N$ and $r_j=\pm 1$
indicate whether the particles move right or left. The function
$A_{\sigma_{Q_1},\ldots,\sigma_{Q_N}}(r_1 k_{p_1},\ldots,r_N
k_{p_N})$ is given as
\begin{eqnarray}
& &A_{\sigma_{Q_1},\ldots,\sigma_{Q_N}}(r_1 k_{p_1},\ldots,r_N
k_{p_N}) \nonumber \\&=&\sum_{q,\epsilon_{\alpha}=\pm
1}S(\epsilon_{q_1}\Lambda_{q_1},\ldots,\epsilon_{q_M}\Lambda_{q_M})\nonumber \\
& & \times \prod_{\alpha=1}^M F_{r_1 k_{p_1},\ldots r_N
k_{p_N}}(\epsilon_{q_{\alpha}} \Lambda_{q_{\alpha}};m_{\alpha}),
\end{eqnarray}
with $m_{\alpha}$ being the indices of the $M$ down spins in an
increasing order $(m_{\alpha}< m_{\alpha+1})$, $q$ denotes the
permutations of $1,2,\ldots,M$, and $\sum_q$ means the summation
over all the permutations. The functions $F$ are
\begin{eqnarray}
F_{k_1,\ldots k_N}(\Lambda;m)=\prod_{j=1}^{m-1}\frac{k_j-\Lambda+i
c/2}{k_j-\Lambda-i c/2}\frac{1}{k_m-\Lambda-i c/2},
\end{eqnarray}
while the coefficients $S$ are determined by the equations
\begin{eqnarray}
\frac{S(\ldots,\Lambda_{\alpha},\Lambda_{\beta},\ldots)}{S(\ldots,\Lambda_{\beta},\Lambda_{\alpha},\ldots)}=
\frac{\Lambda_{\beta}-\Lambda_{\alpha}+i
c}{\Lambda_{\beta}-\Lambda_{\alpha}-i c}.
\end{eqnarray}

The hard wall boundary condition sets the requirement of the wave
function as,
\begin{equation}
\Psi(0\sigma_1,\cdots,x_N\sigma_N)=\Psi(x_1\sigma_1,\cdots,L\sigma_N)=0.
\end{equation}
The eigenvalue equation (3) can then be solved by the coordinate
Bethe ansatz method \cite{CNYang}. The eigenenergy is given by
\begin{equation}
E=\sum_{j=1}^N k_j^2,
\end{equation}
where the quasimomenta $k_j$ should satisfy the Bethe ansatz
equations \cite{F.Woynarovich,NOelkers}
\begin{eqnarray}
k_j L&=&\pi I_j-\sum_{\beta=1}^M \left(
\arctan\frac{k_j-\Lambda_{\beta}}{c/2}+\arctan\frac{k_j+\Lambda_{\beta}}{c/2}\right),
\nonumber\\
&&\sum_{j=1}^N\left(\arctan\frac{\Lambda_{\alpha} -k_j}{c/2}
+\arctan\frac{\Lambda_{\alpha}+k_j}{c/2}\right)=\pi
J_{\alpha} \nonumber \\
&+& \sum_{\beta=1}^M
\left(\arctan\frac{\Lambda_{\alpha}-\Lambda_{\beta}}{c}+\arctan\frac{\Lambda_{\alpha}+\Lambda_{\beta}}{c}\right).
\end{eqnarray}
Here $j=1,2,\ldots,N$ and $\alpha=1,2,\ldots,M$, where $M$ is the
number of the spin-down fermions. The two sets of variables
$\{k_j\}$ and $\{\Lambda_{\alpha}\}$ are the quasi-momenta and
spin rapidities, respectively. $\{I_j\}$ and $\{J_{\alpha}\}$ are
quantum numbers. For the ground state, the quantum number
configuration is described by a continuous sequence:
\begin{eqnarray*}
     I_j&=&j,         1\leq j \leq N;\\
     J_{\alpha}&=&\alpha,       1 \leq \alpha \leq M.
\end{eqnarray*}

The values of quasi-momenta and the explicit form of the wave
function are determined by the solutions of the Bethe ansatz
equations (9). In the case where the interactions among the atoms
are repulsive, we have two interesting limiting regimes. One is
two-component free Fermi gas and the other is infinite repulsion
interaction case which is like the case of indistinguishable
spinless fermions.

If $c=0$, the system (1) degenerates to the noninteracting
two-component Fermi gas. The quasi-momentum takes the value of the
real momentum and the fermions occupy the single particle momentum
states according to the Pauli exclusion principle. The wave function
of $N$ noninteracting fermions with $M$ spin-down fermions is given
by the Slater determinant as:
\begin{eqnarray}
\Psi=\det
   \begin{pmatrix}
 \psi_1(x_1\uparrow)   & \ldots  &\ldots& & \psi_1(x_N\uparrow) \\
 \psi_1(x_1\downarrow)  & \ldots &\ldots& & \psi_1(x_N\downarrow)\\
 \psi_2(x_1\uparrow)  & \ldots &\ldots& & \psi_2(x_N\uparrow)  \\
\vdots &\vdots  &\ddots& &\vdots  \\
 \psi_{N-M}(x_1\uparrow)  & \ldots &\ldots& &
 \psi_{N-M}(x_N\uparrow)
   \end{pmatrix}.
\end{eqnarray}
where $\psi_m=\sqrt{\frac{2}{L}}\sin(\frac{m\pi}{L}),
m=1,2,\ldots,N$ is the single-particle eigenstates of a particle in
an infinite potential.

If $c\rightarrow\infty$, the system (1) degenerates to the spinless
fermions. The fermions cannot feel $\delta$ interaction of each
other as the wave function must vanishes when two spinless fermions
touch each other, so the atoms must occupy the single particle
momentum eigenstates with momenta $k_j=\frac{j\pi}{L}$
$(j=1,2,\ldots,N)$. The anti-symmetric wave function for the $N$
spinless fermions is
\begin{eqnarray}
\Psi=\det
   \begin{pmatrix}
 \psi_1(x_1)   & \psi_1(x_2)  &\ldots& & \psi_1(x_N) \\
 \psi_2(x_1)  & \psi_2(x_2) &\ldots& & \psi_2(x_N)  \\
 \psi_3(x_1)  & \psi_3(x_2) &\ldots& & \psi_3(x_N)  \\
\vdots &\vdots  &\ddots& &\vdots  \\
 \psi_N(x_1)  & \psi_N(x_2) &\ldots& & \psi_N(x_N)
   \end{pmatrix}.
\end{eqnarray}

\section{Single-particle distribution}
\label{sec:sum}

\subsection{Single-particle reduced density matrix and off-diagonal long range order}

In quantum mechanics, the density matrix $\rho_N
=|\Psi\rangle\langle\Psi|$ plays an important role. We first
consider the single-particle reduced density matrix
\begin{equation}
\rho_1=Tr_{23\ldots N}|\Psi\rangle\langle\Psi|,
\end{equation}
where the trace means to do integrations over all the position
coordinates and spin indices except one of them as we are
considering the spin-half particle system. In coordinate
representation, it is given by
\begin{eqnarray*}
&&\rho_1(x,x')=\nonumber \\
&&\quad \frac{N\int_0^L
\Psi^*(x,x_2,...,x_N)\Psi(x',x_2,...,x_N)dx_2\ldots dx_N}{\int_0^L
|\Psi(x_1,x_2,...,x_N)|^2dx_1\ldots dx_N}.
\end{eqnarray*}

\begin{figure}[h]
\begin{center}
\includegraphics[scale=0.90]{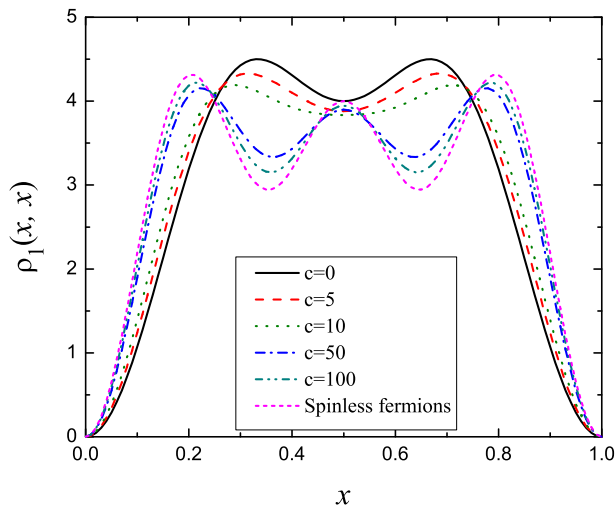}
\caption{\label{fig:epsart2}(color online)  Position density
distributions of fermions with different interactions for
$N=3,M=1$.}
\end{center}
\end{figure}

The diagonal matrix elements $\rho(x,x)$ give the real space
position density distribution of the particles, which are shown in
Fig.~\ref{fig:epsart2}. Starting from noninteracting case $c=0$,
where the fermions occupy the single-particle momentum states, the
position density is the sum of the $N$ independent fermions lying
in their own momentum states according to the Pauli exclusion
principle. When interaction strength increases, the distributions
are smoothened, and the half-width increases gradually. As the
interaction increases further, the position density distribution
tends to that of the spinless fermions.

\begin{figure}[h]
\begin{center}
\includegraphics[scale=1.05]{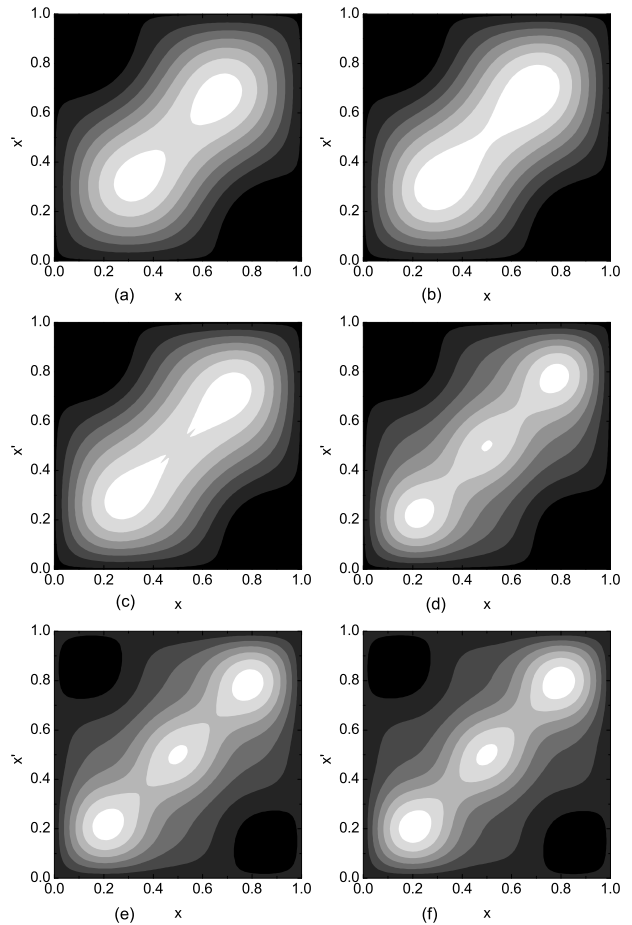}
\caption{\label{fig:epsart1} Single-particle reduced density matrix
$\rho(x,x')$
 as a function of interaction strength for $N=3,M=1$. Shown are the
interactions (a) $c=0$, (b) $c=5$, (c) $c=10$, (d) $c=50$, (e)
$c=100$ and (f) $c=1000$. }
\end{center}
\end{figure}

The off-diagonal matrix elements show the correlations between
particles at different positions which come from quantum
fluctuations. The single-particle reduced density matrix is shown
in Fig.~\ref{fig:epsart1} for different repulsive interactions.
The single-particle reduced density matrix expresses the
self-correlations for a single particle in real space. In
classical mechanics, it is a Dirac delta function $\delta(x-x')$.
 So one can observe from Fig.~1 that a strong enhancement of
$\rho(x,x')$ exists along the diagonal line $x=x'$ for all
interaction strengths.

\subsection{Natural orbitals and their populations}

\begin{figure}[h]
\begin{center}
\includegraphics[scale=0.85]{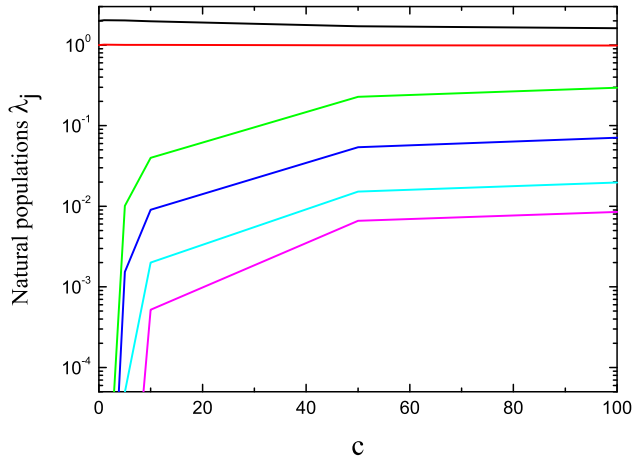}
\caption{\label{fig:epsart3}(color online). Evolution of the
populations of the six lowest natural orbitals $\phi_i$ with the
interactions for the case of $N=3,M=1$.}
\end{center}
\end{figure}
The spectral decomposition of the single-particle reduced density
matrix gives
\begin{equation}
\int dx'\rho_1(x,x')\phi_i(x)=\lambda_i\phi_i(x),  i=1,2,\ldots,
\end{equation}
where $\phi_i(x)$ are the so called natural orbitals, which is
obtained from the eigenfunctions of the single-particle reduced
density matrix and represents an effective single-particle states,
and $\lambda_i$ is population of the $i$-th natural orbital and
satisfies $\sum_i \lambda_i=N$.

We first consider two limit cases. For noninteracting
two-component Fermi gas, the single-particle wave functions
$\psi_i(x)$ are the natural orbitals of the system and
$\lambda_j=2,$ for $j=1,2,...,M$ and $\lambda_j=1$, for
$j=M+1,...,N-M$, all the higher eigenvalues being zero as their
reduced density matrix can be given as
\begin{eqnarray}
\rho_1(x,x')=2\sum_{j=1}^M  \psi_j(x) \psi_j(x')+
\sum_{j=M+1}^{N-M} \psi_j(x) \psi(x'),
\end{eqnarray}
For spinless Fermi gas, the single-particle wave functions
$\psi_i(x)$ are also the natural orbitals of the system and
$\lambda_j=1,$ for $j=1,2,...,N$ and all the higher eigenvalues
being zero,
\begin{equation}
\rho_1(x,x')=\sum_{i=1}^N \psi_i^*(x) \psi_i(x'),
\end{equation}

\begin{figure}[h]
\begin{center}
\includegraphics[scale=1.40]{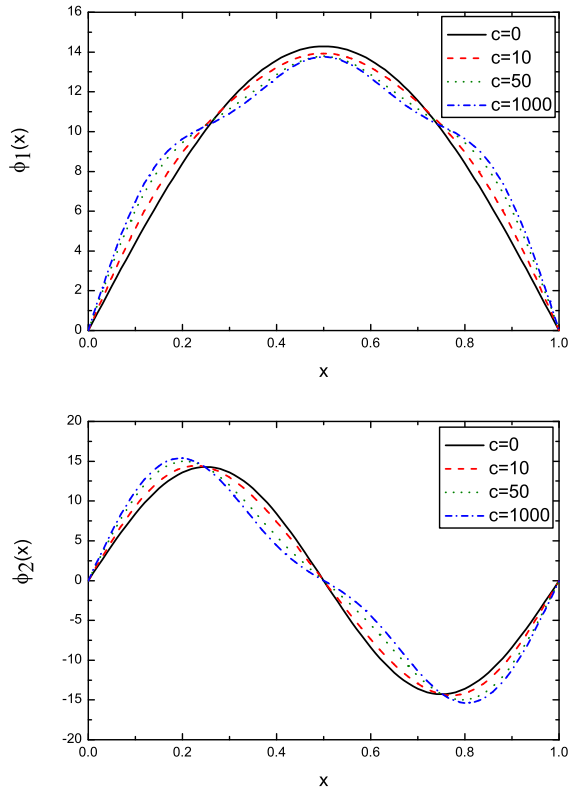}
\caption{\label{fig:epsart4}(color online).  The profile of two
dominant natural orbitals for different interactions for $N=3,M=1$
fermions.}
\end{center}
\end{figure}

For interacting two-component Fermi gas, we find that interaction
leads to significant differences in the spectrum of natural
orbitals and their eigenvalues. The populations of lowest six
natural orbitals $\lambda_i(c)$ with the interactions are shown in
Fig.~\ref{fig:epsart3}. We see that the first two natural orbitals
$(i=1,2)$ are dominant and the other higher orbitals $(i>2)$ are
separated from them. When the interaction increases, the
populations of the lowest two dominant orbitals decreases while
those higher orbitals increase slowly. However, the lowest two
orbitals are still dominant. We show the profiles of the two
dominant natural orbitals for different repulsive interactions in
Fig.~\ref{fig:epsart4}. In the uncorrelated case $c=0$, the system
is in a number state, and natural orbitals coincide with the
single-particle eigenstates. When the interaction increases, the
natural orbitals will be somewhat distorted but the parity of the
orbital is conserved.

\subsection{Momentum density distribution}

\begin{figure}[h]
\begin{center}
\includegraphics[scale=0.95]{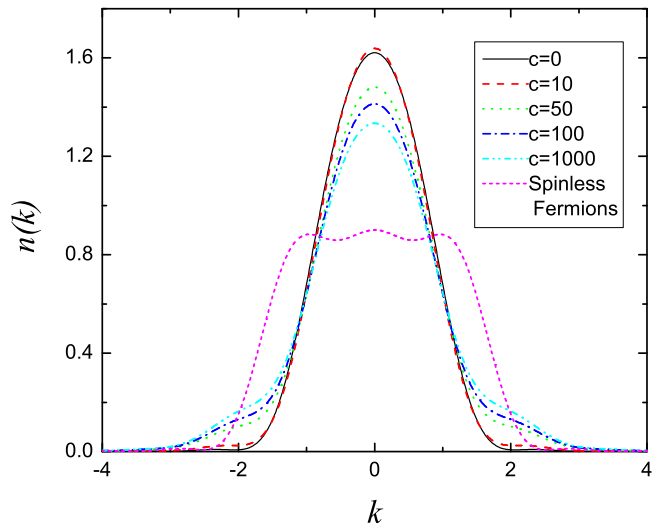}
\caption{\label{fig:epsart5}(color online).  Momentum distribution
of fermions with different repulsive interactions $c$ for
$N=3,M=1$.}
\end{center}
\end{figure}

In experiments, the momentum density distribution detect the
off-diagonal correlations of the single-particle reduced density
matrix. The momentum density distribution is calculated from the
single-particle reduced density matrix
\begin{equation}
n(k)=2\pi^{-1}\int_0^L dx \int_0^L dx' \rho(x,x') e^{-ik(x-x')}.
\label{cc}
\end{equation}
From Eq. (\ref{cc}), we see that the momentum density distribution
$n(k)$ can be understood as the Fourier transform of the
integrated off-diagonal correlation function.
Fig.~\ref{fig:epsart5} shows the evolution of the momentum density
distribution from free fermions to spinless fermions. When the
interaction increases, the distribution becomes smoother, and the
half-width becomes larger. When the interaction $c$ extends to
infinity, we find that the momentum density distribution is still
different from the spinless fermions although they coincide in
real space. This is to be expected as the coordinate exchange
symmetry of the two-component fermions is different from that of
the spinless fermions.

Comparing with the momentum density distribution of a boson gas in
a hard wall trap \cite{HaoA}, we see that the momentum density
distributions for two-component fermions are smoother and the
half-widths are larger. This is due to the Pauli exclusion
principle of fermions, that no more than two fermions can occupy
same momentum state, so in momentum space the density distribution
are broader. For infinite interaction, the hard-core bosons have
the same position density as the spinless fermions, but the
momentum density distributions are different \cite{HaoA,HaoB}. The
reason is that their wave functions have different symmetry, that
is, the wave function of the bosons is symmetric while that of the
fermions is antisymmetric.

\section{two-particle density distribution}

\begin{figure}[h]
\begin{center}
\includegraphics[scale=1.05]{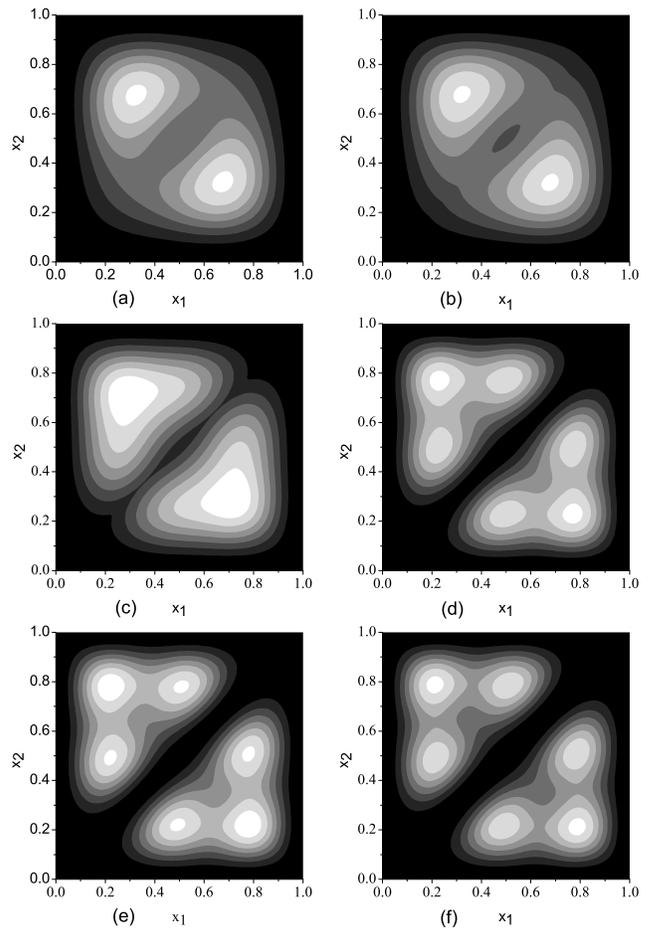}
\caption{\label{fig:epsart6} Two-body density $\rho(x_1,x_2)$ as a
function of interaction strength for three fermions with one
down-spin. Shown are the interactions (a) $c=0$, (b) $c=1$, (c)
$c=10$, (d) $c=50$, (e) $c=100$, and (f) $c=1000$. }
\end{center}
\end{figure}

Now, we consider the two-particle correlations. The two-particle
reduced density matrix is $\rho_2= tr_{34\ldots
N}|\Psi\rangle\langle\Psi|$. In coordinate representation, the
two-particle density, which is the diagonal of the two-particle
reduced density matrix, also called the pair distributions, is
obtained from the many-body wave function $\Psi$ as:
\begin{eqnarray*}
& &\rho_2(x_1,x_2)=\\
&&\frac{N\int_0^L
\Psi^*(x_1,x_2,...,x_N)\Psi(x_1,x_2,...,x_N)dx_3\ldots
dx_N}{\int_0^L |\Psi(x_1,x_2,...,x_N)|^2dx_1\ldots dx_N}.
\end{eqnarray*}
The two-particle density expresses the joint probability of finding
one particle at position $x_1$ and any second one at $x_2$. We show
the two-particle density in Fig.~\ref{fig:epsart6} for different
interactions. We find that the two-particle density is minimized
along the diagonal line for all the interactions due to the Pauli
exclusion principle. When the interaction increases, the
distributions split into two parts, which demonstrates that the
particles are more localized in the well.

\section{Conclusion}
\label{sec:sum}

We have investigated the ground state properties of the
one-dimensional two-component Fermi gas in an infinite potential
well. From the solutions of the Bethe ansatz equations, we
obtained the explicit form of the wave function of the system for
a few atoms. Then we studied the single-particle reduced density
matrix, momentum density distribution and two-particle density
matrix for different interactions. We also found that the momentum
density distributions of the strongly interacting two-component
fermions were very different from that of the spinless fermions,
while the position density distributions of both were the same.

\begin{acknowledgements}

We thanks S. Chen for helpful discussions, and W. L. Chan for the critical
reading of our paper. This work is supported by RGC Grant CUHK 402107, the
National Natural Science Foundation of China, and the National Basic Research
Programme of China.
\end{acknowledgements}


\begin{references}



\bibitem{AGorlitz}
A. G\"{o}rlitz \emph{et al}., Phys. Rev. Lett. \textbf{87}, 130402
(2001).

\bibitem{HMoritz}
H. Moritz, T. St\"{o}ferle, M. K\"{o}hl, and T. Esslinger, Phys.
Rev. Lett. \textbf{91}, 250402 (2003).


\bibitem{TKinoshita}
T. Kinoshita, T. Wenger, and D. S. Weiss, Science \textbf{305},
1125 (2004).

\bibitem{BLTolra}
B. Laburthe Tolra, K. M. O'Hara, J. H. Huckans, W. D. Phillips, S.
L. Rolston, and J. V. Porto, Phys. Rev. Lett. \textbf{92}, 190401
(2004).





\bibitem{RecatiA}
A. Recati, J. N. Fuchs and W. Zwerger, Phys. Rev. A \textbf{71},
033630 (2005).

\bibitem{Tokatly}
I. V. Tokatly, Phys. Rev. Lett. \textbf{93}, 090405 (2004).

\bibitem{MOlshanii}
M. Olshanii, Phys. Rev. Lett. \textbf{81}, 938 (1998).

\bibitem{BEGranger}
B. E. Granger and D. Blume, Phys. Rev. Lett. \textbf{92},
133202(2004); K. Kanjilal and D. Blume, Phys. Rev. A
\textbf{70},042709(2004).



\bibitem{DSPetrov}
D. S. Petrov, G. V. Shlyapnikov, and J. T. M. Walraven, Phys. Rev.
Lett. \textbf{85}, 3745 (2000).


\bibitem{VDunjko}
V. Dunjko, V. Lorent, and M. Olshanii,  Phys. Rev. Lett.
\textbf{86}, 5413 (2001).


\bibitem{POhberg}
P. \"{O}hberg, and L. Santos, Phys. Rev. Lett. \textbf{89}, 240402
(2002).

\bibitem{T.Bergeman}
T. Bergeman, M. G. Moore, and M. Olshanii, Phys. Rev. Lett.
\textbf{91}, 163201 (2003).






\bibitem{HaoA}
Y. J. Hao, Y. B. Zhang, J. Q. Liang, and S. Chen, Phys. Rev. A
\textbf{73}, 063617 (2006).



\bibitem{MTBatchelor}
M. T. Batchelor, X. W. Guan, N. Oelkers, and C. Lee, J. Phys. A
\textbf{38}, 7787 (2005).



\bibitem{MGaudin}
M. Gaudin, Phys. Lett. A \textbf{{\bf 24}}, 55 (1967).


\bibitem{CNYang}
C. N. Yang, Phys. Rev. Lett.  \textbf{{\bf 19}}, 1312 (1967).


\bibitem{F.Woynarovich}
F. Woynarovich, Phys. Lett. A. \textbf{{\bf 108}}, 401 (1985).


\bibitem{NOelkers}
N. Oelkers, M. T. Batchelor, M. Bortz and X. W. Guan, J. Phys. A
\textbf{{\bf 39}}, 1073 (2006).




\bibitem{Tujimotok}
K. Tsujimoto  \emph{et al}, Phys. Lett. A \textbf{277}, 123
(2000); T. Shirai \emph{ et al}, Surf. Sci. \textbf{514}, 95
(2002) .



\bibitem{E. Lieb}
E. Lieb and F. Y. Wu, Phys. Rev. Lett. \textbf{{\bf 20}}, 1445
(1968).





\bibitem{HaoB}
Y. J. Hao, Y. B. Zhang, and S. Chen, Phys. Rev. A \textbf{76},
063601 (2007).


\end{references}
\end{document}